\documentclass[11pt,twoside]{article}
\usepackage{asp2010}
\usepackage{graphicx,wrapfig}

\resetcounters

\newcommand{\msun}{\ensuremath{M_{\odot}}}
\newcommand{\nue}{\ensuremath{\nu_{e}}}
\newcommand{\nuebar}{\ensuremath{\bar \nu_e}}
\newcommand{\numt}{\ensuremath{\nu_{\mu\tau}}}
\newcommand{\numtbar}{\ensuremath{\bar \nu_{\mu\tau}}}
\newcommand{\numu}{\ensuremath{\nu_{\mu}}}
\newcommand{\nutau}{\ensuremath{\nu_{\tau}}}
\newcommand{\numubar}{\ensuremath{\bar \nu_{\mu}}}
\newcommand{\nutaubar}{\ensuremath{\bar \nu_{\tau}}}
\newcommand{\mev}{\mbox{MeV}}
\newcommand{\gcc}{\ensuremath{{\mbox{g~cm}}^{-3}}}
\newcommand{\isotope}[2]{$^{#2}$#1}
\newcommand{\degree}{\ensuremath{^\circ}}

\begin{document}

\title{A Three-Dimensional Neutrino-Driven Core Collapse Supernova Explosion of a 15 M$_\odot$ Star}
\author{
Anthony Mezzacappa$^{1,2}$, 
Eric J. Lentz$^{1,2,3}$, 
Stephen W. Bruenn$^4$,  
W. Raphael Hix$^{1,3}$, 
O.E. Bronson Messer$^{3,5}$, 
Eirik Endeve$^{1,6}$, 
John M. Blondin$^{7}$, 
J. Austin Harris$^{1}$, 
Pedro Marronetti$^{8}$,
Konstantin N. Yakunin$^{2}$,  
and Eric J. Lingerfelt$^{6}$
\affil{
$^1$Department of Physics and Astronomy, University of Tennessee, Knoxville, TN 37996\\
$^2$Joint Institute for Computational Sciences, Oak Ridge National Laboratory, Oak Ridge, TN, 37831\\
$^3$Physics Division, Oak Ridge National Laboratory, Oak Ridge, TN 37831\\
$^4$Department of Physics, Florida Atlantic University, Boca Raton, FL 33431\\
$^5$National Center for Computational Sciences, Oak Ridge National Laboratory, Oak Ridge, TN 37831\\
$^6$Computer Science and Mathematics Division, Oak Ridge National Laboratory, Oak Ridge, TN 37831\\
$^7$Department of Physics, North Carolina State University, Raleigh, NC 27695\\
$^8$Physics Division, National Science Foundation, Arlington, VA 22230
}
}

\begin{abstract}
We present results from an {\em ab initio} three-dimensional, multi-physics core collapse supernova simulation for the case of a 15 M$_\odot$ progenitor. Our simulation includes multi-frequency neutrino transport with state-of-the-art neutrino interactions in the ``ray-by-ray'' approximation, and approximate general relativity. Our model exhibits a neutrino-driven explosion. The shock radius begins an outward trajectory at approximately 275 ms after bounce, giving the first indication of a developing explosion in the model. The onset of this shock expansion is delayed relative to our two-dimensional counterpart model, which begins at approximately 200 ms after core bounce. At a time of 441 ms after bounce, the angle-averaged shock radius in our three-dimensional model has reached 751 km. Further quantitative analysis of the outcomes in this model must await further development of the post-bounce dynamics and a simulation that will extend well beyond 1 s after stellar core bounce, based on the results for the same progenitor in the context of our two-dimensional, counterpart model. This more complete analysis will determine whether or not the explosion is robust and whether or not observables such as the explosion energy, $^{56}$Ni mass, etc. are in agreement with observations. Nonetheless, the onset of explosion in our {\em ab initio} three-dimensional multi-physics model with multifrequency neutrino transport and general relativity is encouraging.
\end{abstract}

\section{Introduction}

Numerous investigations over the past several decades clearly indicate the need to model core collapse supernovae in three spatial dimensions with realistic neutrino transport and general relativity [for a review, see \cite{KoSaTa06,Janka12,MeBrLe15}]. Realistic neutrino transport requires at a minimum a spectral, or multifrequency, treatment, where the lowest-order angular moments of the neutrino distribution function are evolved -- e.g., the energy- and momentum-density per frequency. The models are therefore phase-space, and not purely spatial, models -- i.e., at a minimum, carried out in the four-dimensional space of neutrino frequency and three spatial dimensions and, eventually, in the full, six-dimensional space of neutrino directions cosines, energy, and three spatial dimensions. Either of these approaches ("moments" or full Boltzmann) will require supercomputer architectures at the 100 PF -- 1 EF scale, and perhaps beyond. Furthermore, the challenge to include a realistic treatment of neutrino transport {\em and} general relativity adds significant complexity to both of these model components, of course, but especially to the neutrino transport \citep{CaMe03,CaEnMe13a,CaEnMe13b}. Consequently, in an effort to advance to three-spatial dimensions {\em with} necessary physics -- e.g., multifrequency neutrino transport and general relativity -- one compromise has been developed and is used here: the ray-by-ray transport approximation \citep{buraja03}. In this approximation, a set of spherically symmetric transport problems are solved for each ``ray." A ray is an outgoing radial ray for each $(\theta,\phi)$ pair. Note, neutrino transport in this case is not restricted to pure radial transport. Rather, a complete spherical symmetric solve is performed. The {\em net} neutrino flux is however purely radial in this approximation. The more spherically symmetric the central source -- in this case the proto-neutron star -- the more valid the approximation. A quantitative assessment of the quality of the approximation awaits an analysis using both ray-by-ray and non-ray-by-ray approaches in the context of an otherwise identical model. In the meantime, the ray-by-ray approach affords an ability to explore three-dimensions with the neutrino transport sophistication of past spherically symmetric models, at the expense of lateral neutrino transport.

Recently, results from three-dimensional multi-physics models with detailed ray-by-ray neutrino transport and approximate general relativity were reported by the ``Oak Ridge'' and Max Planck groups \citep{MeBrLe15,MeJaMa15}. Here, we focus on the results from our ongoing model. We report on the phases through the initiation of explosion. Additional analyses -- e.g., a determination of the explosion energy -- await the completion of 1--1.5 s of postbounce evolution.

\section{The CHIMERA Code}

CHIMERA is a parallel, multi-physics code built specifically for multidimensional simulation of core collapse supernovae.
It is the chimeric combination of separate codes for hydrodynamics and gravity; neutrino transport and opacities; and a nuclear EoS and reaction network, coupled by a layer that oversees data management, parallelism, I/O, and control.
The hydrodynamics are modeled using a dimensionally-split, Lagrangian-Remap (PPMLR) scheme  \citep{CoWo84} as implemented in VH1 \citep{HaBlLi12}.
Self-gravity is computed by multipole expansion \citep{MuSt95}. In the GR case, the Newtonian monopole is replaced with a GR monopole  \citep[][Case~A]{MaDiJa06}.
Neutrino transport is computed in the ``ray-by-ray-plus'' (RbR+) approximation \citep{buraja03}.
Neutrinos are advected laterally (in the $\theta$ and $\phi$ directions) with the fluid and contribute to the lateral pressure gradient where $\rho>10^{12}\,\gcc$.
The transport solver is an improved and updated version of the multi-group flux-limited diffusion transport solver of \citet{Brue85} enhanced for GR \citep{BrDeMe01}, with an additional geometric flux limiter to prevent the over-rapid transition to free streaming of the standard flux-limiter.  All $O(v/c)$ observer corrections have been included.
CHIMERA solves for all three flavors of neutrinos and antineutrinos with four coupled species: \nue, \nuebar, $\numt=\{\numu,\nutau\}$, $\numtbar=\{\numubar,\nutaubar\}$, with typically 20 energy groups each for $\alpha\epsilon =  4-250~\mev$, where $\alpha$ is the lapse function and $\epsilon$ is the comoving-frame group center neutrino energy.
Our standard, modernized, neutrino--matter interactions include emission, absorption, and non-isoenergetic scattering on free nucleons \citep{reprla98}, with weak magnetism corrections \citep{Horo02}; emission/absorption (electron capture) on nuclei \citep{lamasa03}; isoenergetic scattering on nuclei, including ion-ion correlations; non-isoenergetic scattering on electrons and positrons; and pair emission from $e^+e^-$-annihilation \citep{Brue85} and nucleon-nucleon bremsstrahlung \citep{hara98}.
CHIMERA generally utilizes the $K = 220$~\mev\ incompressibility version of the \citet{lasw91} EoS for  $\rho>10^{11}\,\gcc$ and a modified version of the \citet{Coop85} EoS for  $\rho<10^{11}\,\gcc$, where nuclear statistical equilibrium (NSE) applies.
Most CHIMERA simulations have used a 14-species $\alpha$-network ($\alpha$, \isotope{C}{12}-\isotope{Zn}{60}) for the non-NSE regions \citep[XNet][]{HiTh99a}.
To aid the transition between the network and NSE regimes, we have constructed a 17-species NSE solver to be used in place of the Cooperstein EoS for electron fractions $Y_{\rm e}> 0.46$.
An extended version of the Cooperstein electron--photon EoS is used throughout.

During evolution the radial zones are gradually and automatically repositioned during the remap step to track changes in the radial structure.
To minimize restrictions on the time step from the Courant limit, the lateral hydrodynamics for a few inner zones are ``frozen'' during collapse, and after prompt convection fades the laterally frozen region expands to the inner 8--10~km.
In the ``frozen'' region, lateral velocities are set to 0 and the lateral hydrodynamic sweep is skipped. The full radial hydrodynamics and neutrino transport are always computed to the center of the simulation for all rays.

The simulation presented here utilizes 32,400 rays (solid angle elements) with 2\degree\ resolution in longitude and a resolution in latitude that varies from 8\degree\ at the pole to better than 0.7\degree\ at the equator, but is uniform in the cosine of the colatitude. 
Due to the Courant limit, the coordinate pole in standard spherical-polar coordinates creates a strong restriction on the time step size and therefore lengthens the total run time compared to a similar resolution two-dimensional simulation.  
Our constant cosine-of-colatitude grid seeks to minimize this impact without resorting to a grid that is coarse at all latitudes or implementing unevolved (frozen) regions near the pole. 

\section{Results}\label{sec:current3D}

\begin{figure}
\includegraphics[angle=270,width=4.0in]{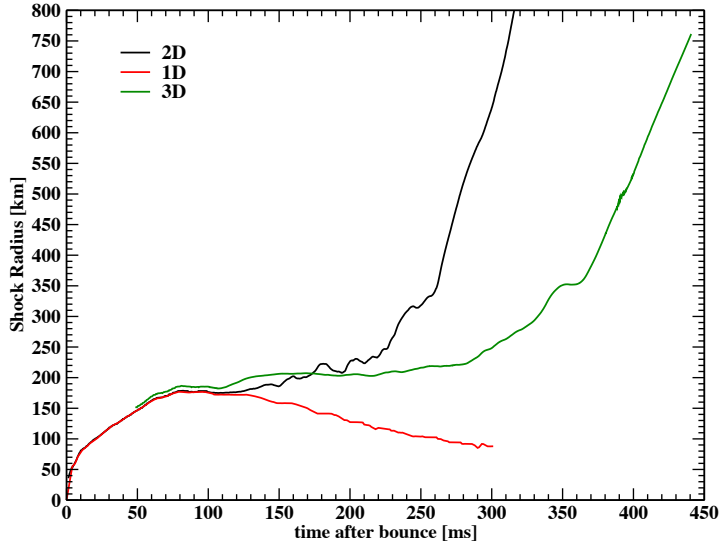}
\caption{Evolution of the shock trajectory from our 1D model and the angle-averaged shock trajectories from our two- and three-dimensional models, all for the 15~\msun\ case \citep{LeBrHi15}. The 1D model does not develop an explosion, whereas an explosion is obtained in both our two-dimensional and our three-dimensional models.
\label{fig:1D2D3DShockTrajectories}
}
\end{figure}

\begin{figure}
\includegraphics[angle=90,width=4.0in]{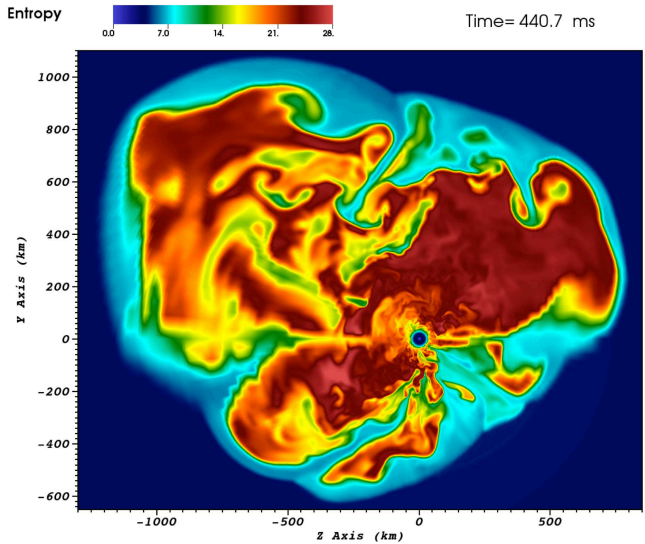}
\caption{Snapshot of the equatorial cross section of the entropy in our ongoing three-dimensional simulation for the 15~\msun\ case at $\sim$441 ms after bounce \citep{LeBrHi15}. Red indicates high-entropy, expanding, rising material. Green/blue indicates cooler, denser material. Evident are significant (green) down flows fueling the neutrino luminosities.
\label{fig:entropy3D}
}
\end{figure}

\begin{figure}
\includegraphics[width=4.0in]{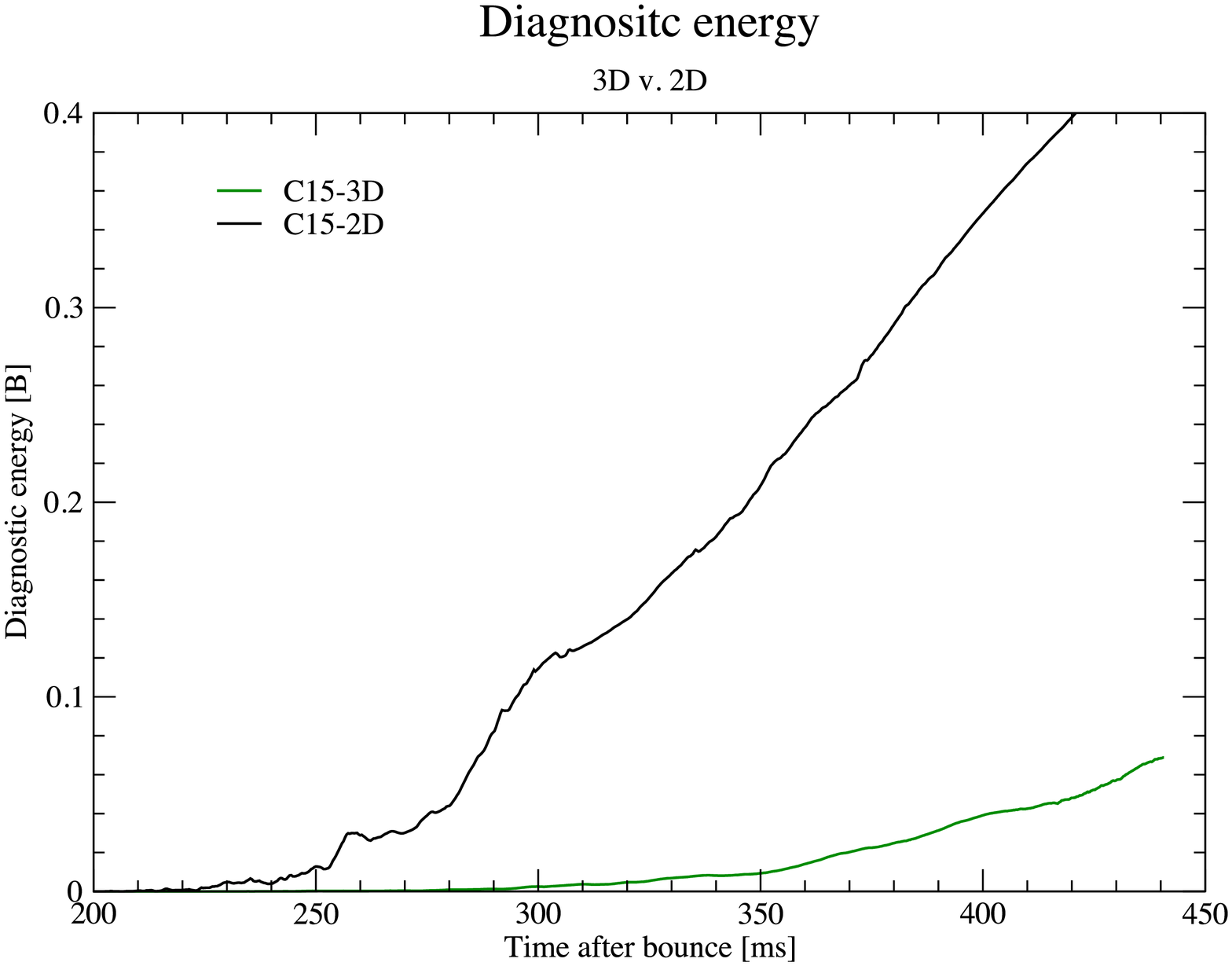}
\caption{\label{fig:diag}
The diagnostic energy as a function of postbounce time for our ongoing three-dimensional model and its two-dimensional counterpart \citep{LeBrHi15}.
}
\end{figure}

In Figure \ref{fig:1D2D3DShockTrajectories} we plot the angle-averaged shock radii as a function of time after bounce for our one-, two-, and three-dimensional models. All three models are initiated from the same 15 M$_\odot$ progenitor, taken from the most recent Woosley--Heger series \citep{WoHe07}, and include the complete physics set described in the previous section. After an initial leveling off of the shock radius at $\sim$ 180 km between 75--100 ms after bounce, the outcomes begin to diverge, with the one-dimensional shock radius peeling off first and moving inward in a monotonic fashion until this particular simulation is halted. The two- and three-dimensional shock radii, however, maintain a value between 180 and 220 km between 100 ms and $\sim$ 190 ms after bounce, at which point the radii computed in the two- and three-dimensional models begin to diverge, with the shock radius in the two-dimensional case rising rapidly and the three-dimensional radius remaining at $\sim$ 200--210 km until it too begins to rise rapidly at $\sim$ 280 ms after bounce.

Figure \ref{fig:entropy3D} provides a snapshot of our ongoing three-dimensional model at 441 ms after bounce. Specifically, we show an equatorial slice through the postshock volume, of the stellar core entropy. The shock wave is outlined by the jump in entropy across it. Neutrino-driven convection is evident in the pattern of hot (red) rising plumes, which bring neutrino-heated material up to the shock, and cool (green) down flows, which replace the fluid below. Clearly the distortion of the shock and the patterns of convection beneath it break axismmetry. Evidence for $l=1$, ``sloshing'' and $m=1$, ``spiral'' modes of the SASI will require a modal analysis, although the two-dimensional slice does not rule out either mode. Of particular note, at this instant of time one massive accretion funnel extends down to the proto-neutron star. Such funnels are a natural outcome of SASI-channeled postshock flows. Their importance lies in the fact they continue to fuel the neutrino luminosities on long time scales, of order 1 s after bounce.

Figure \ref{fig:diag} shows the evolution of the diagnostic energy in our three-dimensional model. Also plotted, for comparison, is the diagnostic energy for its counterpart two-dimensional model. The diagnostic energy is the sum of the total energy (gravitational plus internal plus kinetic) for all zones on our numerical grid for which this total is positive. The diagnostic energy does not take into consideration the work required to lift the stellar material above it, nor the energy gained in nuclear recombination as the initially outgoing material continues to move outward, potentially becoming unbound. With both of these included, we could define the explosion energy, but at this time after bounce, the explosion energy in our model is still not positive. However, the diagnostic energy does in fact become positive at approximately 300 ms after bounce in our three-dimensional model. In the three-dimensional case, this is achieved at a time that is delayed relative to the time at which it is achieved in the two-dimensional case, by approximately 50 ms.

\section{Conclusions}

Within the context of the approximations made, the results presented here indicate that the neutrino-driven explosions obtained in the context of two-dimensional core collapse supernova simulations also obtain here, in the context of a three-dimensional model. While the approximations made -- most notably, the ray-by-ray neutrino transport approximation, as well as the general relativistic monopole correction to the Newtonian gravitational potential -- must be assessed quantitatively in the context of future models that can accurately compare and contrast such approaches, we believe our use of detailed, spherically symmetric transport along each ray and a general relativistic monopole correction to the Newtonian potential in the context of the three-dimensional model presented marks a significant step forward. Clearly, an additional $\sim$1 second of postbounce evolution is required in the model presented, and numerous models for different progenitors, using different equations of state, etc. must be developed, as well. For these, we must have significant patience. Models such as the one presented here will run for a significant fraction of a year on either the DOE Oak Ridge Leadership Computing Facility or NSF's Blue Waters platform. Thus, core collapse supernova modeling is entering a new phase, requiring the utilization of national-class resources for extended periods. As a result, the number of models that can be produced in a given year will be limited, and each model will require a significant time to complete.

\section{Acknowledgements}

We acknowledge support from the DOE Office of Nuclear Physics and Office of Advanced Scientific Computing Research, NASA's Astrophysics Theory Program (grant no. NNH11AQ72I), and the NSF PetaApps Program (grant no. OCI-0749242). We also acknowledge compute resources on Jaguar and Titan at the Oak Ridge Leadership Computing Facility and on Hopper at the National Energy Research Scientific Computing Center, through DOE's Innovative and Novel Computational Impact on Theory and Experiment (INCITE) Program, and on Kraken, at the National Institute for Computational Sciences, through NSF's XRAC Program.


\begin{thebibliography}{}
\expandafter\ifx\csname natexlab\endcsname\relax\def\natexlab#1{#1}\fi
\expandafter\ifx\csname url\endcsname\relax
  \def\url#1{\texttt{#1}}\fi
\expandafter\ifx\csname urlprefix\endcsname\relax\def\urlprefix{URL }\fi
\providecommand{\eprint}[2][]{\url{#2}}

\bibitem[{Bruenn(1985)}]{Brue85}
Bruenn, S.~W. 1985, ApJS, 58, 771

\bibitem[{{Bruenn} et~al.(2001){Bruenn}, {De Nisco}, \&
  {Mezzacappa}}]{BrDeMe01}
{Bruenn}, S.~W., {De Nisco}, K.~R., \& {Mezzacappa}, A. 2001, ApJ, 560, 326

\bibitem[{{Buras} et~al.(2003){Buras}, {Rampp}, {Janka}, \&
  {Kifonidis}}]{buraja03}
{Buras}, R., {Rampp}, M., {Janka}, H.-T., \& {Kifonidis}, K. 2003, Phys. Rev.
  Lett., 90, 241101

\bibitem[{{Cardall} et~al.(2013{\natexlab{a}}){Cardall}, {Endeve}, \&
  {Mezzacappa}}]{CaEnMe13b}
{Cardall}, C.~Y., {Endeve}, E., \& {Mezzacappa}, A. 2013{\natexlab{a}}, \prd,
  88, 023011. \eprint{1305.0037}

\bibitem[{{Cardall} et~al.(2013{\natexlab{b}}){Cardall}, {Endeve}, \&
  {Mezzacappa}}]{CaEnMe13a}
--- 2013{\natexlab{b}}, \prd, 87, 103004. \eprint{1209.2151}

\bibitem[{{Cardall} \& {Mezzacappa}(2003)}]{CaMe03}
{Cardall}, C.~Y., \& {Mezzacappa}, A. 2003, Phys. Rev. D, 68, 023006

\bibitem[{{Colella} \& {Woodward}(1984)}]{CoWo84}
{Colella}, P., \& {Woodward}, P. 1984, J. Comp. Phys., 54, 174

\bibitem[{{Cooperstein}(1985)}]{Coop85}
{Cooperstein}, J. 1985, Nucl. Phys. A, 438, 722

\bibitem[{{Hannestad} \& {Raffelt}(1998)}]{hara98}
{Hannestad}, S., \& {Raffelt}, G. 1998, ApJ, 507, 339

\bibitem[{{Hawley} et~al.(2012){Hawley}, {Blondin}, {Lindahl}, \&
  {Lufkin}}]{HaBlLi12}
{Hawley}, J., {Blondin}, J., {Lindahl}, G., \& {Lufkin}, E. 2012, Astrophysics
  Source Code Library, 4007

\bibitem[{{Hix} \& {Thielemann}(1999)}]{HiTh99a}
{Hix}, W.~R., \& {Thielemann}, F.-K. 1999, ApJ, 511, 862

\bibitem[{{Horowitz}(2002)}]{Horo02}
{Horowitz}, C.~J. 2002, Phys. Rev. D, 65, 43001

\bibitem[{{Janka}(2012)}]{Janka12}
{Janka}, H.-T. 2012, Annual Review of Nuclear and Particle Science, 62, 407.
  \eprint{1206.2503}

\bibitem[{{Kotake} et~al.(2006){Kotake}, {Sato}, \& {Takahashi}}]{KoSaTa06}
{Kotake}, K., {Sato}, K., \& {Takahashi}, K. 2006, Reports on Progress in
  Physics, 69, 971. \eprint{arXiv:astro-ph/0509456}

\bibitem[{{Langanke} et~al.(2003){Langanke}, {Mart{\'\i}nez-Pinedo}, {Sampaio},
  {Dean}, {Hix}, {Messer}, {Mezzacappa}, {Liebend{\"o}rfer}, {Janka}, \&
  {Rampp}}]{lamasa03}
{Langanke}, K., {Mart{\'\i}nez-Pinedo}, G., {Sampaio}, J.~M., {Dean}, D.~J.,
  {Hix}, W.~R., {Messer}, O.~E., {Mezzacappa}, A., {Liebend{\"o}rfer}, M.,
  {Janka}, H.-T., \& {Rampp}, M. 2003, Phys. Rev. Lett., 90, 241102

\bibitem[{Lattimer \& Swesty(1991)}]{lasw91}
Lattimer, J., \& Swesty, F.~D. 1991, Nucl. Phys. A, 535, 331

\bibitem[{{Lentz} et~al.(2015){Lentz}, {Bruenn}, {Hix}, {Mezzacappa}, {Messer},
  {Endeve}, {Blondin}, {Harris}, {Marronetti}, \& {Yakunin}}]{LeBrHi15}
{Lentz}, E.~J., {Bruenn}, S.~W., {Hix}, W.~R., {Mezzacappa}, A., {Messer},
  O.~E.~B., {Endeve}, E., {Blondin}, J.~M., {Harris}, J.~A., {Marronetti}, P.,
  \& {Yakunin}, K.~N. 2015, ArXiv e-prints.
  \eprint{1505.05110}

\bibitem[{{Marek} et~al.(2006){Marek}, {Dimmelmeier}, {Janka}, {M{\"u}ller}, \&
  {Buras}}]{MaDiJa06}
{Marek}, A., {Dimmelmeier}, H., {Janka}, H.-T., {M{\"u}ller}, E., \& {Buras},
  R. 2006, A\&A, 445, 273. \eprint{arXiv:astro-ph/0502161}

\bibitem[{{Melson} et~al.(2015){Melson}, {Janka}, \& {Marek}}]{MeJaMa15}
{Melson}, T., {Janka}, H.-T., \& {Marek}, A. 2015, ArXiv e-prints.
  \eprint{1501.01961}

\bibitem[{{Mezzacappa} et~al.(2015){Mezzacappa}, {Bruenn}, {Lentz}, {Hix},
  {Harris}, {Bronson Messer}, {Endeve}, {Chertkow}, {Blondin}, {Marronetti}, \&
  {Yakunin}}]{MeBrLe15}
{Mezzacappa}, A., {Bruenn}, S.~W., {Lentz}, E.~J., {Hix}, W.~R., {Harris},
  J.~A., {Bronson Messer}, O.~E., {Endeve}, E., {Chertkow}, M.~A., {Blondin},
  J.~M., {Marronetti}, P., \& {Yakunin}, K.~N. 2015, ArXiv e-prints.
  \eprint{1501.01688}

\bibitem[{M{\"u}ller \& Steinmetz(1995)}]{MuSt95}
M{\"u}ller, E., \& Steinmetz, M. 1995, Comp. Phys. Comm., 89, 45

\bibitem[{{Reddy} et~al.(1998){Reddy}, {Prakash}, \& {Lattimer}}]{reprla98}
{Reddy}, S., {Prakash}, M., \& {Lattimer}, J.~M. 1998, Phys. Rev. D, 58, 013009

\bibitem[{{Woosley} \& {Heger}(2007)}]{WoHe07}
{Woosley}, S.~E., \& {Heger}, A. 2007, Phys. Rep., 442, 269.
  \eprint{arXiv:astro-ph/0702176}

\end{thebibliography}
\end{document}